\documentstyle[12pt]{article}

\font\tenrm=cmr10
\font\tenit=cmti10
\font\elevenbf=cmbx10 scaled\magstep 1
\font\elevenrm=cmr10 scaled\magstep 1
\font\elevenit=cmti10 scaled\magstep 1

\font\ninerm=cmr9

\input epsf

\font\zfont = cmss10 scaled \magstep1
\def\ZZ{\hbox{\zfont Z\kern-.4emZ}}
\def\ie{{\elevenit i.e.}}

\def\Tr{{\rm Tr}}
\def\beq{\begin{equation}}
\def\eeq{\end{equation}}
\def\beqra{\begin{eqnarray}}
\def\eeqra{\end{eqnarray}}
\def\refmark#1{\cite{#1}}
\def\lsim{{<}}
\def\gsim{{>}}
\def\fnote#1#2{\begingroup\def\thefootnote{#1}\footnote{#2}\addtocounter
{footnote}{-1}\endgroup}

\def\bibj#1#2{{\bibit #1~}{\bibbf #2}}
\def\prl{Phys. Rev. Lett.}
\def\prd{Phys. Rev.}
\def\npb{Nucl. Phys.}
\def\plb{Phys. Lett.}

\newcommand{\bibit}{\elevenit}
\newcommand{\bibbf}{\elevenbf}
\renewenvironment{thebibliography}[1]
 { \elevenrm
   \begin{list}{\arabic{enumi}.}
    {\usecounter{enumi} \setlength{\parsep}{0pt}
     \setlength{\itemsep}{3pt} \settowidth{\labelwidth}{#1.}
     \sloppy
    }}{\end{list}}
\renewcommand{\thefootnote}{\fnsymbol{footnote}}
\begin{document}
\hfill SCIPP/93-14

\hfill June 1993
\vglue 1.0cm

\begin{center}
\vglue 0.6cm
{{\elevenbf THE STRONG CP PROBLEM, STRING THEORY\\}
 {\elevenbf        \vglue 10pt
AND THE NELSON-BARR MECHANISM\\}}

\vglue 1.0cm
{\tenrm ROBERT G. LEIGH\fnote{\dagger}{\ninerm\baselineskip=11pt Address after
September 1, 1993: Department of Physics and Astronomy,
Rutgers University, Piscataway, NJ 08855-0849.}\\}
\baselineskip=13pt\vglue 3pt
{\tenit Santa Cruz Institute for Particle Physics\\}
\baselineskip=12pt
{\tenit University of California, Santa Cruz, CA 95064\\}
\vglue 2.0cm
{\tenrm ABSTRACT \vglue 5pt}
\end{center}

\vbox{\narrower
 \tenrm\baselineskip=12pt
 \noindent
We review recent work on the strong CP problem in the context of
realistic string-inspired models. We discuss the various solutions,
review the conjecture that CP is generally a gauged discrete symmetry
in string theory and then consider models of the Nelson-Barr type.
We note that squark non-degeneracy
spoils the Nelson-Barr structure at the one loop level.
We stress that string theory expectations, as well as naturalness
arguments, make it very difficult to avoid the constraints on
non-degeneracy.}

\vglue 0.9cm
\vfill
\vskip4.5cm
\vbox{
\centerline{\it to appear in the proceedings of
Recent Advances in the Superworld,}
\centerline{\it Houston Advanced Research Center,
The Woodlands, TX, April 1993}
 }
\vskip1cm
\pagebreak
\baselineskip=14pt
\elevenrm

The strong CP problem is well known. In a CP non-invariant theory
there is an additional operator, proportional to a parameter $\theta$,
which leads to a large contribution to the neutron electric dipole
moment. By comparison to the experimental data, it is found that
$\theta$ must be less than $\sim 10^{-9}$. There are two contributions
to $\theta$: a bare value, present in theories with explicit CP violation,
and a contribution proportional to the overall phase of the mass matrix
of the coloured fields, $\theta_{QFD} \sim arg\; Det\; m$. That
there is no reason for these two contributions to combine to leave such
a small remainder constitutes the strong CP problem.

There are three well known solutions to the strong CP problem. The
first scenario simply has no observable $\theta$-parameter. This
arises, for example, when $m_u=0$. This solution however is manifestly
inconsistent with lowest
order chiral perturbation theory,\refmark{muzero} which
gives $m_u/m_d \simeq 0.5$. Recently however it has been suggested
that an ambiguity at next-to-leading order may allow a zero up-quark
mass. This subject is at best controversial, and we will refrain
from commenting further on it here, except to note that such
a solution to the strong CP problem would be relatively easy to
implement; the required symmetry structure is very simple.

The second solution to the strong CP problem has been discussed
extensively. Given a spontaneously broken anomalous symmetry (the
Peccei-Quinn symmetry),
$\theta$ may be dynamically relaxed to zero through the vacuum
expectation value of the axion.\refmark{pq,axion}
The axion solution is tightly constrained by astrophysical
and cosmological considerations.\refmark{axioncosmology}
In the context of string
theory, there are many potential sources of axions.\refmark{lopnan}
It is well known that the gravity supermultiplet gives rise to a
model-independent axion, which couples universally to all of the
gauge groups of a given four-dimensional model:
\beqra  *da = H=dB-\omega_Y+\omega_L  \nonumber \\
\Delta a = *dH = *(R\wedge R-F\wedge F) .\eeqra
The problem with this axion is that its decay constant is of order
the string scale, much larger than that required by the usual
cosmological arguments.

Other axions may arise from other sources, depending upon the
particular compactification studied, such as from internal components
of the antisymmetric tensor field. However, the associated PQ symmetries
are generally violated\refmark{dine1} by worldsheet non-perturbative
effects and thus the associated axions are inappropriate for the
strong CP problem.

In the presence of an `anomalous' $U(1)$ in the
spectrum of a string model, there is an interesting effect on the
model independent axion. The anomaly is cancelled by the transformation
of a Green-Schwarz term:
\beq \delta \Gamma_{GS} \propto \int B\wedge F \rightarrow \int A\wedge *da .
\eeq
Thus the associated gauge boson, $A$, gains mass by eating the model
independent axion. The local $U(1)$ symmetry is therefore broken and
the model-independent axion is no longer present in the spectrum.
However, there is a remaining {\elevenit global} $U(1)$ symmetry, which
is spontaneously broken. There is always an associated Fayet-Iliopoulos
$D$-term\refmark{dine2} present; in general there exists a supersymmetric
vacuum in which the $D$-term is cancelled by vev's of some charged
scalar fields. This breaking of the global symmetry leads to a new axion
with decay constant of order the scalar vev; perhaps, it may thus be
possible to attain a decay constant at much smaller scales.

Even if an appropriate axion can be found, there are additional problems.
First, if other gauge groups become strong at a high scale, the axion
potential will be dominated by instanton effects of this gauge group
and the
QCD $\theta$ angle will not generically be cancelled. This will happen
for example, in models in which gaugino condensation occurs.
Essentially, one axion is necessary for each strong gauge group.

In addition to this problem, it is widely believed that
quantum gravitational effects, such as from black holes or wormholes,
lead to violations\refmark{jmr} of global symmetries (such as
PQ symmetries) through higher dimensional operators whose scale
is set by the Planck mass.
Such operators give direct contributions to the axion potential and
lead to a minimum away from $\theta=0$ unless they can be
suppressed through
some very high dimension. It is known\refmark{casasross} that
this may occur accidentally in some models: an array of gauge and discrete
symmetries can be constructed such that an approximate PQ symmetry
exists which ensures a sufficiently small $\theta$.
This mechanism is perhaps the most likely axion scenario within the
context of string theory, though we note the rather delicate symmetry
structure involved.

Given the problems outlined above, it is natural to consider the third
solution to the strong CP problem, the Nelson-Barr mechanism.\refmark{nelbarr}
In this scenario, one supposes that CP is a good symmetry of the
underlying theory, but is spontaneously violated.
In this case, the bare value of $\theta$ is zero and one arranges things
so that the tree level mass matrix is complex but possesses a real
determinant. The game is then to race against perturbative corrections
to keep $\theta$ small enough. The scale of CP violation should be low
enough that higher dimensional operators not spoil the structure of the
theory; we will take this to be roughly of order $10^{11}$ GeV, a scale
which can be naturally generated in string models (given $m_{susy}\sim$ TeV).

It is this scenario that we will explore in the context of string-inspired
models in this paper. To begin however, we will make a few remarks
concerning CP as a spontaneously violated symmetry.

\vglue 0.6cm
{\elevenbf\noindent 1. CP as a Spontaneously Violated Symmetry}
\vglue 0.2cm

It is not at all obvious {\elevenit a priori} that the Nelson-Barr
mechanism can be implemented in string theory. We must first argue that
CP {\elevenit can be} a spontaneously violated symmetry.
Until recently it was quite possible that CP was
simply not a good symmetry (it could for example have been violated by
non-perturbative effects). However, it turns out that CP is an example
of a {\elevenit gauged} discrete symmetry. Because of this it is protected
from all manner of violation, whether non-perturbative stringy phenomena
or otherwise (such as quantum gravitational effects). The fact that
it is seen to be violated in our low energy world must then be a consequence
of spontaneous breakdown, perhaps through complex vev's of heavy scalar
fields.

In fact it is possible to demonstrate\refmark{cppaper} the gauge symmetry
that gives rise to the four-dimensional CP symmetry in a wide range of
simple string compactifications. In ten-dimensions, the heterotic
string theory possesses no parity symmetry as the theory is chiral. There
is however a charge conjugation symmetry which can easily be seen to
be equivalent to an $SO(32)$ (or $E_8\times E_8$) gauge transformation.
In the simplest toroidal
compactifications, the four dimensional theory is non-chiral; a parity
symmetry can be constructed which has just the right transformation
properties on the states. It is a proper (ten-dimensional) Lorentz
transformation. Four-dimensional charge conjugation is a combination
of the aforementioned gauge rotation and a proper Lorentz transformation.

Thus in these simplest of theories, both C and P can be thought
of as gauge symmetries: they arise as combinations of ordinary gauge
and general cooordinate transformations. For suitable values of moduli,
many other compactifications possess a gauged CP symmetry. It is reasonable to
conjecture, as a general property of string theory, that CP is
indeed a gauge symmetry.

If this is true, CP can be violated by complex expectation values
of fields. In the remainder of this paper we will suppose that CP is
spontaneously violated by complex expectation values of observable sector
matter fields at a relatively low
scale ($\sim 10^{11}$ GeV). In many models, CP might be violated by
complex vev's of hidden sector fields or moduli. There is little hope
that this could give rise to a successful implementation of the
Nelson-Barr mechanism, and we will simply assume that this is not
realized in the models of interest. We will see that Nelson-Barr
models quite generically have severe problems; the origin of these
problems is in the soft supersymmetry breaking (supergravity) sector.

\vglue 0.6cm
{\elevenbf\noindent 2. One Loop contributions to $\theta$}
\vglue 0.2cm

In supersymmetric models of spontaneous CP violation where one has arranged
the vanishing of $\theta$ at tree-level, there are a variety of
possible contributions to $\theta$ at one-loop
order.  These are given by
\beq\delta \theta=Im\; \Tr\;\left[ m_u^{-1} \delta m_u + m_d^{-1}
\delta m_d\right]-3
{\rm Arg} \;{\tilde m}_3~. \label{eq:delth}\eeq
$\delta m_{u,d}$ represent the one-loop corrections to the tree level
quark mass matrices, $m_{u,d}$, and ${\tilde m}_3$ is the gluino mass including
one-loop contributions. The dominant contributions at one-loop will
come from gluino-squark or quark-squark exchange, respectively. In order to
analyze these quantities, we first need to make a few
more stipulations about the underlying model. To obtain
vanishing $\theta$ at tree level\refmark{barrmasiero,dkl}
the tree level gluino mass must be real (a non-zero phase represents a
contribution to $\theta$). All terms in
the Higgs potential must also be real.
To accomplish this we assume, as discussed above, that
supersymmetry breaking dynamics do not
spontaneously break CP, so that at some large scale the theory is
completely CP-invariant and, in particular, all soft supersymmetry breaking
terms are real.

The supersymmetric Nelson-Barr model that we will consider is
defined as follows: in addition to the usual quark and
lepton families, we have an additional pair of isosinglet down quark
fields, $q$ and $\bar q$, as well as some
singlet fields, ${\cal N}_i$ and $\bar {\cal N}_i$.
It is straightforward to consider models with several $q$ and $\bar q$
fields, and with additional types of singlets.
The terms in the superpotential which give rise to the quark mass
matrix are\footnote{$Q \bar q$ terms can be
forbidden via either gauged $U(1)$ or discrete symmetries.}
\beq W= \mu q \bar q + \gamma^{ij} {\cal N}_i q \bar d_j
+ H_1\lambda_{ij} Q_i \bar d_j\label{eq:basicw}\eeq
(the terms in the superpotential involving $u$ quarks and leptons
will not be important for our considerations).
I should mention here that one framework
for obtaining such a mass matrix is suggested by $E_6$ models,
such as those which often appear in superstring theories.\refmark{frampton}
In these models, generations of quarks and
leptons arise from the {\bf 27} representation, as could the $q$ and
$\bar q$ fields (as members of the {\bf 10} of $SO(10)$).

Given the couplings above, the fermion mass matrix has the
Nelson-Barr structure:
\beq m_F= \left( \matrix{m_d & M_D \vec a \cr 0 &\mu} \right),
\label{eq:fermionmatrix}\eeq
where $m_d=\lambda_d H_1$ and the vector $\vec a$ is defined as:
\beq a^i ={1 \over M_D} \gamma^{ji}{\cal N}_j
{}~~~~~;~~~~~\vec a^{\dagger} \vec a=1.\label{eq:vectora}\eeq
If the ${\cal N}$ fields have complex vev's, CP is spontaneously
broken, but nevertheless the determinant of $m_F$ is
real. In these expressions, $\mu$ and $M_D$ are of order the
intermediate scale.\footnote{$\mu$ may itself be proportional to the
{\elevenit real} vev of some other scalar field. Plausible
mechanisms have been suggested for obtaining such vev's.
In particular, in ``intermediate scale scenarios," it has been noted
that the ${\cal N}$ fields can readily obtain vev's of order
$m_I= \sqrt{m_{3/2} m_{Pl}}$.} We note that in the limit in which we
integrate out the heavy modes, the phases in $\vec a$ lead
to an unsuppressed CKM phase (as long as $\mu\sim M_D$).
Since the dominant one-loop graphs involve squark exchange,
the form of the scalar mass matrices is of particular interest.
Consider first the $\phi \phi^*$ type terms.  For the
squarks in the {\bf 3} representation of $SU(3)$
, these take the form, on the full $4\times 4$ set of states:
\beq {\cal M}_{LL}^2 =
 \left(\matrix{\tilde m_d^2 + m_d^T m_d &
M_D m_d^T \vec a \cr \vec a^{\dagger}m_d M_D &
\tilde m_q^2 + M_D^2 + \mu^2}\right).\label{eq:dsquark}\eeq
Similarly, for the $\bar{\bf 3}$ squarks, we have:
\beq {\cal M}_{RR}^2 =
 \left(\matrix{\tilde m_{\bar d}^2 + m_d m_d^T + M_D^2
\vec a \vec a^{\dagger} &
\mu M_D \vec a \cr \mu M_D \vec a^{\dagger}
& \tilde m_{\bar q}^2 + \mu^2}\right) .\label{eq:dbarsquarks}\eeq
Finally, for the $\phi\phi$-type matrix, which connects the $3$
and $\bar 3$ squarks, we have
\beq {\cal M}_{RL} ^2 = \left(\matrix{A_d <H_1 > + \mu_H{H_2\over H_1} m_d
& M_5^2 \vec b \cr 0 & A_{\mu} \mu} \right)
\label{eq:lrmatrix}\eeq
where $\mu_H$ is the real coefficient of the $H_1H_2$ term in
the superpotential, and $M_5$ and $\vec b$ are defined by
\beq M_5^2\; b^i= A_\gamma^{ji}{\cal N}_j + \left(
{\partial W \over \partial {\cal N}_j}\right)^*
\gamma^{ji}~~~~ ; ~~~~\vec b^{\dagger}
\vec b=1\label{eq:bdefinition}\eeq
Note that $\vec b$ receives contributions from both soft terms as well
as $F$-terms and in general $\vec b$ is not proportional to $\vec a$.
In fact $\vec b$ is of the utmost importance to our discussion. From
the form of the mass matrices given above, it can be seen that there is
mixing of light with heavy states that will lead to effects
unsuppressed by inverse powers of the heavy mass scale
\beq \sim (m_{susy} \mu)\;\; {1\over\mu^2}\;\; (m_{susy} \mu) \eeq
unless a detailed constraint is satisfied. This constraint is
\beq M_5^2 \;\vec b = A_\mu\; M_D\; \vec a .\label{eq:decoupleh}\eeq
We will see the disastrous effects of this potential non-decoupling
in the following.

In a general supergravity setting, the matrices $\tilde m_d^2$,
$\tilde m_{\bar d}^2$ and $A_d$ are completely arbitrary. It
is well-known that flavor changing neutral current processes
(predominantly $K^o-\bar K^o$ mixing) constrain these to be
approximately degenerate, \ie
\beqra \tilde m_{\bar d}^2 = \tilde m_{\bar d}^2 \times {\bf 1}+\delta
m_{\bar d}^2 ~ ; ~~\tilde m_Q^2 = \tilde m_d^2 \times {\bf 1} +\delta
\tilde m_d^2. \nonumber \\
\tilde m_{\bar q}^2 = \tilde m_{\bar d}^2 + \delta
\tilde m_{\bar q}^2~ ; ~~
\tilde m_q^2 = \tilde m_d^2 + \delta\tilde m_q^2
\label{eq:fourbyfourdegen}\eeqra
with $\delta m^2$ small\footnote{To be more precise, the $sd$
components are bounded at the $10^{-2}$ or $10^{-3}$ level by
$K^o-\bar K^o$.} compared
to $m_{susy}^2$. We will find in the following discussion that the
constraints on these quantities from $\theta$ are {\elevenit much}
stronger than from FCNC's.

Let us now ask how severe are the constraints arising from the
smallness of $\theta$.  First consider the gluino mass diagram
shown in Fig. 1.
\vglue 6pt
%
%
%
\vglue 12pt
\begin{center}
\leavevmode\epsfysize=7pc  \epsfbox{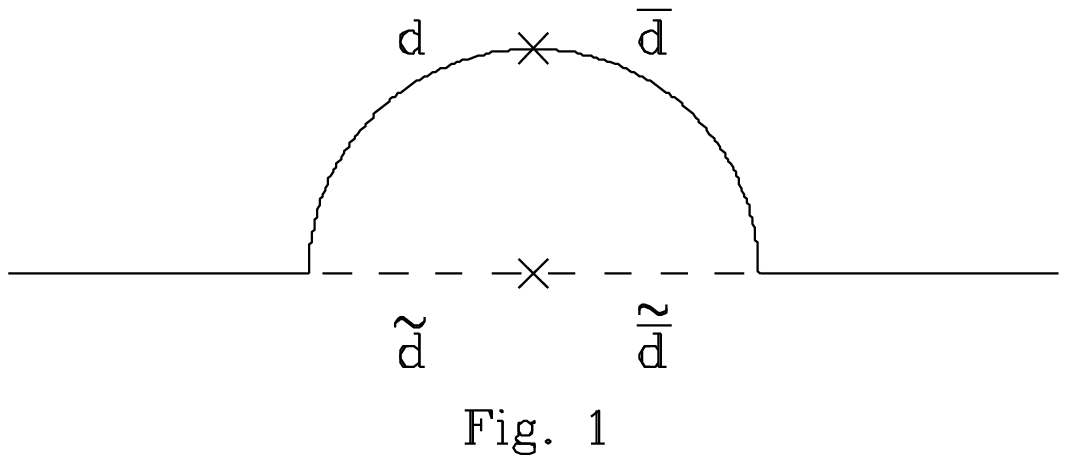}
\end{center}
If the heavy eigenstates are not decoupled (see
discussion above) then $\theta$ gets a potentially
large contribution\refmark{dkl} from:
\beq {\rm Im}~ \delta m_{\lambda}\sim
 {\alpha_s \over 4 \pi}
{M_D^2 \over (M_D^2+\mu^2)} {M_5^2\over M_D}
{\rm Im}~ \vec a^{\dagger} \vec b.\label{eq:imlambda}\eeq
Indeed, this diagram leads to the requirement that the phases of the
vectors $\vec a$ and $\vec b$ line up to about one part in $10^{-7}$.
Certainly the simplest
way to satisfy this is $\vec b =\vec a$; we will assume
this to be the case in the remainder of this section.
Later we will investigate this condition and argue that
it is not natural.

The light fermion contributions to the gluino mass
lead to a  weaker limit on proportionality.  If $A_d$ is not
proportional to the unit matrix, one will obtain a complex
result, in general.  This will give a limit suppressed by
powers of the $b$-quark mass over the susy-breaking scale:
\beq {{\rm Im}~{\vec a}^{\dagger}\delta A_d  m_d^{T}{\vec a}
\over m_{susy}^3} \;{<H_1> M_D^2 \over{ M_D^2 +\mu^2}}\; \lsim\;
 10^{-7}. \label{eq:gluinoalimit}\eeq

More significant limits arise from the graph of Fig. 2.  From
one proportionality violating insertion and one degeneracy
violating insertion we obtain:
\beq {{\rm Im}~{\vec a}^{\dagger}  \delta A_d \lambda_d^{-1}
\left(\delta\tilde m^2_{\bar d}-
\delta \tilde m^2_{\bar q}\times
{\bf 1}\right) \vec a  \over m^3_{susy}}\;{M_D^2 \over {M_D^2 + \mu ^2}}
\;\lsim\; 10^{-6},\label{eq:deltaadeltam}\eeq
and
\beq {{\rm Im}~ {\vec a}^{\dagger} \delta A_d \lambda_d^{-1} {\vec a}\over
m_{susy}^3}\; {(M_5^2 -A_{\mu} M_D)^2 \mu^2 \over{ (M_D^2 + \mu^2)^2}}
\;\lsim\; 10^{-6}. \label{eq:deltaadeltamlh}\eeq

%
%
%
\vglue 12pt
\begin{center}
\leavevmode\epsfysize=7pc \epsfbox{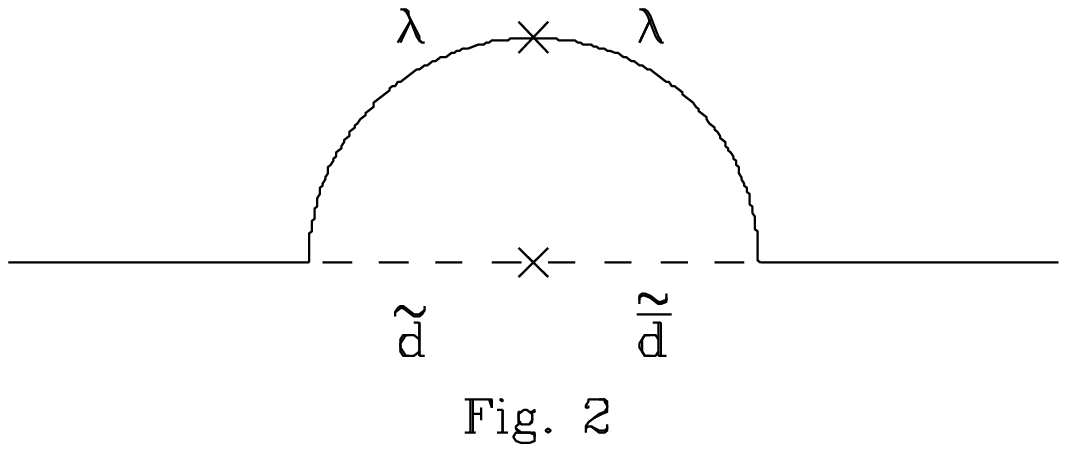}
\end{center}

{}From Fig. 2, with two degeneracy-violating
insertions, we find contributions to $\theta$ of order:
\beq 10^{-1}{\alpha_s\over4\pi}\;{A_d {\rm Im}~\vec a^{\dagger}
m_d \delta\tilde m_Q^2 m_d^{-1} \left(\delta \tilde m^2_{\bar d}-
\delta\tilde m^2_{\bar q}\times{\bf 1}\right)
\vec a  \over m_{susy}^5} {M_D^2 \over {M_D^2 +
\mu^2}},\label{eq:twodeginsertions}\eeq
which lead to the constraints
\beq {\left( \delta \tilde m_{\bar q}^2,
 \delta \tilde m_{\bar d}^2\right)\over m_{susy}^2}
{(\delta \tilde m_Q^2) \over m_{susy}^2} \;\lsim\;
10^{-9}.\label{eq:finallimits}\eeq
Non-degeneracies also lead to a contribution of order
\beq 10^{-1} {\alpha_s\over4\pi}\;{ {\rm Im}~\vec a^{\dagger} m_d \delta\tilde
m_d^2 m_d^{-1}\vec a  \over m_{susy}^3
}{{(M_5^2 -A_{\mu} M_D)M_D} \over {M_D^2 + \mu^2}}.
 \label{eq:twodeginsertionslh}\eeq
We note that Eqs. (\ref{eq:deltaadeltamlh}) and
(\ref{eq:twodeginsertionslh}) vanish in the limit in which
Eq. (\ref{eq:decoupleh}) is satisfied but Eqs.
(\ref{eq:deltaadeltam}) and (\ref{eq:twodeginsertions}) represent
large contributions independent of this.

\vglue 0.6cm
{\elevenbf\noindent 3. Discussion}
\vglue 0.2cm

Given these bounds on parameters, we must ask how natural it is
to expect that they are satisfied. The strongest constraint comes
from $\vec b\neq \vec a$. One notes from the definitions of these
vectors that they are apparently unrelated. There is one
possibility: that the minimum of the vacuum energy occurs at such
values of the fields so that (see Eq. \ref{eq:bdefinition})
\beq {\partial W \over \partial{\cal N}_i} \propto m_{susy} {\cal N}_i^* .\eeq
{}From a careful study of this minimization problem, one can show that
such relations can only hold within a supergravity theory with minimal
K\"{a}hler potential: any non-degeneracy at greater than a part in
$10^7$, such as in the mass matrix of the ${\cal N}$ fields,
will show up as $\theta>10^{-9}$. The degeneracy and proportionality
constraints require satisfying limits on $\delta \tilde m_d^2$,
$\delta \tilde m_{\bar d}^2$, $\delta
\tilde m_{\bar q}^2$ and $\delta A_d$, which are considerably more
stringent than those obtained from $K^o$-$\bar K^o$ mixing.  It is
hard to comprehend how they could be satisfied in the absence of a
detailed theory of flavor. These constraints require as well
a condition on $M_5$, or perhaps some other condition on parameters.
For example, the equality $M_5^2=M_D A_{\mu}$ would eliminate
light-heavy squark couplings, so that contributions to $\theta$
which arise from integrating out heavy fields would vanish; see Eqs.
(\ref{eq:deltaadeltamlh}) and (\ref{eq:twodeginsertionslh}) and the
discussion above. However, this condition requires exact degeneracy
(at the Planck scale).

To see what this means, recall the supergravity induced potential:
\beq V= e^{K}[({\partial W \over \partial \phi^i} +
d_i W)g^{i \bar j}
({\partial W \over \partial \phi^j} + d_{j} W)^* - 3 \vert W \vert^2]
. \label{eq:generalpotential}\eeq
with
\beq d_i = {\partial K \over \partial \phi^{i}}
{}~~~~~;~~~~~g_{i \bar j}= {\partial^2 K \over \partial \phi^i \partial
\phi^{j*}}.\label{eq:kahlermetric}\eeq

If we assume that supersymmetry is broken in a hidden sector,
there are two sets of fields: $z_i$, responsible for
supersymmetry breaking, and the ``visible sector fields," $y_i$, with
\beq W=g(y) + h(z).\label{eq:visibleandhidden}\eeq
Universality\refmark{hlw} is the assumption that there is an approximate
$U(n)$ symmetry of the K\"ahler potential, $K$, where $n$ is the number
of chiral multiplets in the theory.  Frequently one takes simply
\beq K= \sum \phi_i^* \phi_{ i} .\label{eq:flatmetric}\eeq
Clearly, this is an extremely strong assumption.  The Yukawa couplings
of the theory exhibit no such symmetry.  It does
not hold, for example, for a generic superstring
compactification, where the symmetry violations are
simply ${\cal O}(1)$. It is these violations of universality that
lead to non-degeneracy and non-proportionality in the scalar mass
matrices.
We can characterize the violations of universality quite precisely.
For small $y$, we can expand $K$ in powers of $y$.  Rescaling
the fields, we can write
\beq K= k(z, z^*)+ y_i y_i^* +  \ell_{ij}(z,z^*) y_i y_j^*
+ h_{ij} (z,z^*) y_i y_j + ....\label{eq:kahler}\eeq
There is no reason, in general, why $\ell_{ij}$ should
be proportional to the unit matrix, so
the $zz^*$ components of the metric will contain
terms involving $y_i y_j^* $ which are non-universal.
Plugging into Eq. (\ref{eq:generalpotential}) yields non-universal
mass terms for the visible sector fields.  In general, there
is no symmetry which can forbid these couplings; for example,
$\ell = z^* z$ cannot be eliminated by symmetries.
Violations of proportionality arise in a similar manner.

We can discuss the likelihood of enjoying the required
universality on two fronts. Recently the
soft supersymmetry breaking sector that might arise in string theory
has been analyzed.\refmark{vadimjan} It is found that if the
dilaton $F$-term dominates
supersymmetry breaking, then the supergravity is minimal, with a K\"{a}hler
potential given by Eq. (\ref{eq:flatmetric}). However, the analysis is
valid at string tree level only; non-universal corrections are expected
at order $\alpha_{str}/\pi$.  Furthermore, in models of gaugino condensation,
this does not happen anyway; supersymmetry breaking is instead dominated
by F-terms of moduli. Since, unlike the dilaton, moduli are
non-universally coupled the K\"{a}hler potential is in principle the most
general real
functional allowed by gauge invariance, as in Eq. (\ref{eq:kahler}).
In either case, it is certainly true that the high degrees of degeneracy  and
proportionality necessary for the Nelson-Barr mechanism are very
difficult to understand  (at least without some detailed theory
of flavor).

Apart from string theory, we can give general arguments based on
naturalness (in the sense of 't Hooft\refmark{thooft}) for the expected size of
non-degeneracy. In Nelson-Barr theories, as in those studied above,
there are typically coloured fields that live in different
representations of the gauge group. Hence we can really not expect that
universality holds (at the Planck scale) to better than the order
of gauge couplings. For fields in the same representation, we
expect universality to be violated by powers of Yukawa couplings.
To avoid this, it would be necessary to have a complicated
arrangement of flavor symmetries. As yet, we have been unable to find
an example.

Let us now put these naturalness and stringy arguments aside
and simply {\elevenit insist} upon minimal supergravity
at the Planck scale; we can then ask whether or not renormalization group
effects lead to a large $\theta$. The answer to this question is that quite
generically, $\theta$ is in fact too large. There is however one case where
this can be avoided: if there exist no gauge symmetries that
distinguish $\bar d$ from $\bar q$, then $\theta$ can be small enough
if we require a rather mild condition on the couplings $\gamma_{ij}$.

Alternatively, one can ask whether or not there is some (small) range
of parameters for which $\theta$ is small enough. One could, for example,
try to exploit the fact that $M_D \ll \mu$ would suppress all of the
above contributions to $\theta$. Unfortunately, this strategy is
limited by the fact that the induced KM phase or the phase entering
the SUSY box graph would be of order ${M_D^2 \over \mu^2}$ and so this
ratio must be $\gsim 10^{-2}$ in order to generate large enough $\epsilon$.

Thus we are left with the strong feeling that the Nelson-Barr mechanism
is very hard to implement in supersymmetric string-inspired models.
If it is to succeed at all, a great deal of knowledge of
flavor symmetries would be necessary. Certainly such symmetry would
be very complicated, and one wonders if such a solution is really
satisfactory at all. The $m_u=0$ solution or even the automatic axion
scenario are likely to be much easier to attain.

I wish to thank Michael Dine and Alex Kagan for a most enjoyable
collaboration and the organizers of this workshop for their
hospitality.

\pagebreak

\begin{center}
\vglue 0.6cm
{\elevenbf References}
\vglue 0.2cm
\end{center}

\end{document}